# Nano-optomechanical Nonlinear Dielectric Metamaterials

Artemios Karvounis[1], Jun-Yu Ou[1], Weiping Wu[1], Kevin F. MacDonald[1], and Nikolay I. Zheludev[1,2]

[1] *Optoelectronics Research Centre & Centre for Photonic Metamaterials,*
*University of Southampton, Southampton, SO17 1BJ, UK*

[2] *The Photonics Institute & Centre for Disruptive Photonic Technologies,*
*Nanyang Technological University, Singapore 637371*

By harnessing the resonant nature of localized electromagnetic modes in a nanostructured silicon membrane, an all-dielectric metamaterial can act as nonlinear medium at optical telecommunications wavelengths. We show that such metamaterials provide extremely large optomechanical nonlinearities, operating at intensities of only a few µW per unit cell and modulation frequencies as high as 152 MHz, thereby offering a path to fast, compact and energy efficient all-optical metadevices.

Non-metallic metamaterial nanostructures currently attract intense attention as they promise to reduce the losses and costs associated with the use of noble metals in traditional plasmonic architectures.[1] It has already been shown that oxides and nitrides,[2] graphene,[3] topological insulators,[4] and high-index dielectrics[5-13] can be used as platforms for the realization of high-$Q$ resonant metamaterials. A variety of non-metallic media, such as graphene,[14] carbon nanotubes,[15] liquid crystals[16] and semiconductors,[17,18] have also been engaged through hybridization with plasmonic metamaterials to create media with strongly enhanced optical nonlinearities, while a nano-optomechanical nonlinearity has recently been observed in a plasmonic metamaterial.[19] Here we experimentally demonstrate an all-dielectric metamaterial, fabricated from a free-standing semiconductor nano-membrane, with sharp near-infrared optical resonances. It exhibits a strong optical nonlinearity associated with light-induced nano-mechanical oscillations of the structure, which change the physical configuration and thus the resonant response of the metamaterial array's constituent metamolecules.

Optical forces at the sub-micron scale can be comparable or even stronger than elastic forces, and resonantly enhanced optical forces in photonic metamaterials have been theoretically studied for both plasmonic and all-dielectric structures.[20-22] The exchange of energy between incident light and a nano-mechanical resonator can be further enhanced when the light is modulated at the mechanical eigenfrequency of the resonator, indeed it has recently been shown that plasmonic metamaterials can be optically reconfigured on this basis with light modulated at MHz frequencies.[19] In consequence of the fact that the mechanical eigenfrequencies of objects are dictated by their stiffness (Young's modulus) and dimensions, nano-scale mechanical oscillators made of silicon offer the prospect of mechanical vibration at hundreds of MHz or even GHz frequencies.

Considerable efforts have been devoted to the reduction of radiative losses in resonant plasmonic metamaterials, as non-radiative losses (Joule heating) are unavoidable in the constituent metals. In 'all-dielectric' metamaterials non-radiative losses are *a-priori* limited, so with appropriate design they can present even stronger optical resonances, and thereby generate stronger optical forces, than plasmonic counterparts.[22] Previous works have demonstrated that high-index media such as silicon can support optical frequency resonances[5-13] and we harness that characteristic here to engineer an ultrathin medium with optical properties that are highly sensitive to structural reconfiguration.

The metamaterial is fabricated by direct focused ion beam (FIB) milling of a commercially sourced (Norcada Inc.), 100 nm thick polycrystalline silicon membrane in a silicon frame (Fig 1). To date, all-dielectric metamaterials have invariably been realized as 'positive' structures – arrays of discrete high-index features (nanorods, discs, bars, etc.) supported on a lower-index substrate.[5-13] The metamaterial employed in the present study shows however that strong localized resonances can also be excited in 'negative' dielectric nanostructures, i.e. a pattern formed by slots cut into a continuous layer of high-index material. The free-standing configuration has the additional advantage of maximizing refractive index contrast with the near-field environment and thereby resonance quality factor.[23] Each 1.05 µm × 1.05 µm unit cell (metamolecule) contains a rectangular nano-cantilever of length $L$ = 300 nm and width $W$ = 600 nm, with an additional slot across the fixed end of the cantilever arm to increase flexibility (Fig 1c shows a geometric schematic of the structure). The metamaterial array is composed of 25 × 25 metamolecules.

This structure supports several optical resonances in the near-infrared range, as illustrated by the microspectrophotometrically measured reflection, transmission and derived absorption spectra presented in Figs. 1d-f. These data show good correlation with spectra obtained via 3D finite element numerical modelling (COMSOL Multiphysics), using a fixed complex refractive index for polycrystalline silicon of 3.2 + 0.04i (following Ref. 24 with an imaginary part tuned to match resonance quality with experimental observations, using the 1550 nm pump wavelength as a reference point: a value ≤0.01 for pristine polycrystalline silicon produces much sharper spectral features – the elevated value employed here effectively represents a variety of material and manufacturing imperfections including deviations from the ideal rectilinear, perfectly planar geometry of the model, surface roughness, and gallium contamination from the FIB milling process).

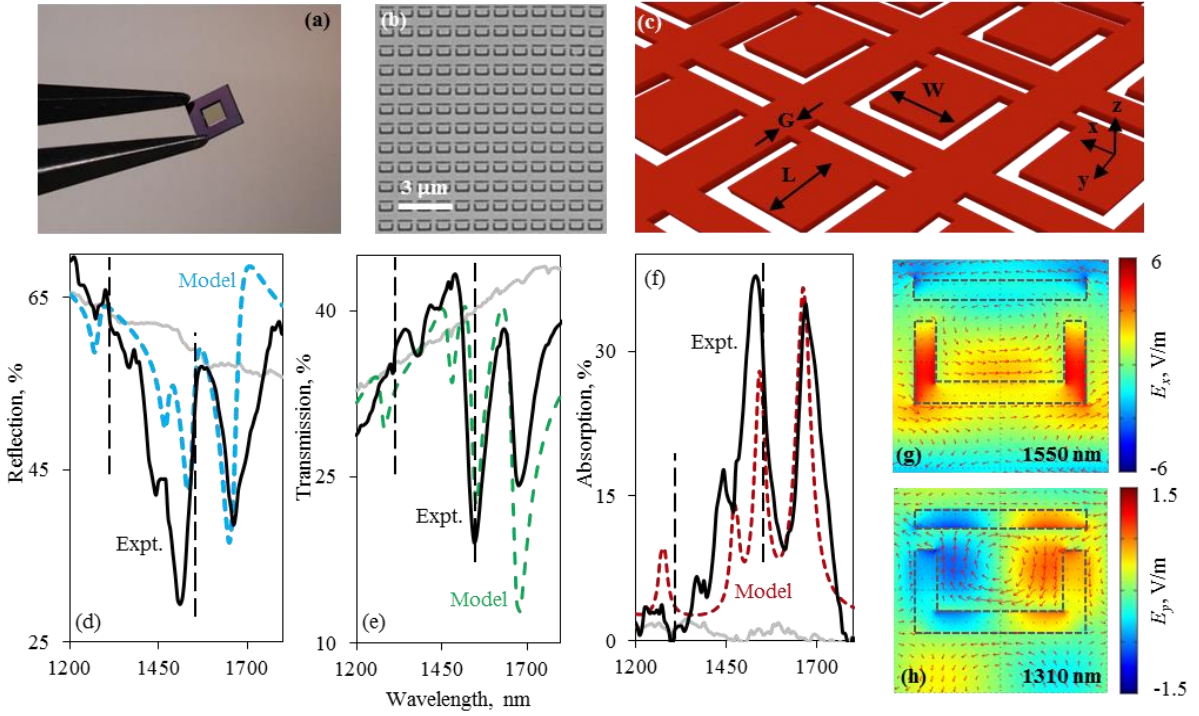

**Figure 1. Optical resonances in free-standing silicon nano-membrane metamaterials:** (a) 100 nm thick Si membrane in 5 mm × 5 mm Si frame used as a platform for fabrication of free-standing all-dielectric metamaterials. (b) Scanning electron microscope image of part of a nano-cantilever metamaterial array fabricated in a Si membrane by focused ion beam milling [dark areas = slots cut through the membrane]. (c) Schematic oblique view of the nano-cantilever array: period = 1.05 μm; $L$ = 300 nm; $W$ = 600 nm; $G$ = 100 nm; slot width = 100 nm. (d-f) Normal incidence reflection, transmission and absorption spectra of the metamaterial for $x$-polarized light. Black lines correspond to experimental measurements, dashed lines to numerical modelling results; Grey lines show measured spectra for the unstructured silicon membrane. (g, h) Field maps, in the $xy$ plane at the mid-point of the membrane thickness, for (g) the electric mode resonance at 1550 nm [$E_x$ field component] and (h) the magnetic mode at 1310 nm [$E_y$ field] – the experimental pump and probe wavelengths respectively. The field maps are overlaid with arrows indicating the direction and magnitude of electric displacement.

In order to achieve a strong optomechanical nonlinearity we require a metamaterial that is highly sensitive at the probe wavelength (1310 nm in the present case) to structural reconfiguration driven by strong optical forces generated within the structure at the pump wavelength (1550 nm). Figure 2a shows simulated transmission around 1310 nm and corresponding maps of field distribution at this wavelength for three different configurations – tilt angles – of the metamolecule nano-cantilevers. The probe wavelength sits to one side of a resonance based upon a spatial distribution of electric field and displacement currents in the cantilever arms that generates magnetic dipoles (in the manner of the familiar plasmonic asymmetric split ring metamaterial 'trapped mode'[25]). The spectral dispersion depends strongly on the cantilever tilt angle, with the resonance blue-shifting as the cantilever arms tilt out of plane (i.e. as their effective length decreases), resulting in a transmission increase at the fixed 1310 nm probe wavelength (there is a concomitant reflectivity decrease, and no meaningful change in absorption).

The absorption resonance in the 1550 nm waveband is derived from the excitation of an electric dipole, hybridized with higher order electric multipoles, within each unit cell (Fig. 1g), giving rise to a spatial distribution of optical forces that tilts the cantilever arms out of the sample plane. In classical electrodynamics the components of the total time-averaged force $F$ acting on an object illuminated with light can be calculated using the surface integral:[26]

$$\langle F_i \rangle = \oiint_S \langle T_{ij} \rangle n_j dS \qquad (1)$$

$S$ being a closed surface around the region of interest, $n_j$ unit vector components pointing out of the surface and $\langle T_{ij} \rangle$ the time-averaged Maxwell stress tensor defined by

$$\langle T_{ij} \rangle = \tfrac{1}{2} Re \left[ \varepsilon \varepsilon_0 \left( E_i E_j^* - \tfrac{1}{2} \delta_{ij} |E|^2 \right) + \mu \mu_0 \left( H_i H_j^* - \tfrac{1}{2} \delta_{ij} |H|^2 \right) \right] \qquad (2)$$

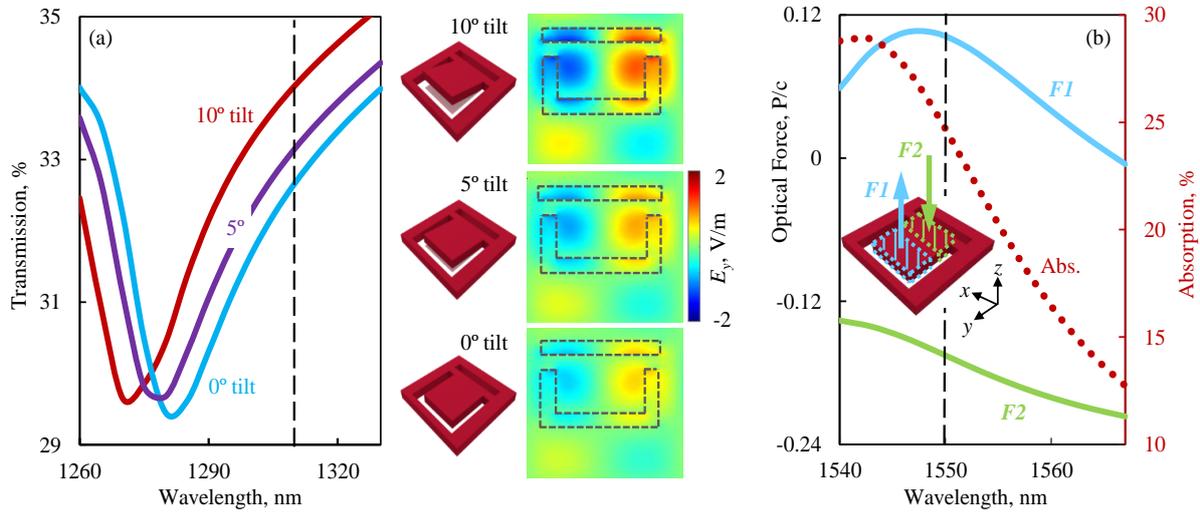

**Figure 2. Modelling optical forces and changes in optical properties resulting from structural reconfiguration:** (a) Numerically simulated dispersion of metamaterial transmission around the experimental 1310 nm probe wavelength for a selection of nano-cantilever tilt angles, as labelled and illustrated schematically to the right, alongside corresponding $E_y$ field maps for the xy plane at the mid-point of the membrane thickness at 1310nm. (b) Spectral dispersion of the normalized out-of-plane optical forces acting on either end of the metamolecule nano-cantilevers [as illustrated inset], and of metamaterial absorption, around the experimental 1550 nm pump wavelength. [In all cases, simulations assume normally-incident, x-polarized light.]

The optical force given by Eq. (1) encompasses both radiation pressure, which arises through the transfer of momentum between photons and any object on which they impinge, and the gradient force, which is associated with intensity variations in the local field around an object. Applied to the metamaterial unit cell (Fig. 2b), this stress tensor analysis reveals antiparallel forces acting on the two ends of the cantilever arms – a net 'positive' force *F1* (in the +z direction towards the light source) at the free end of the arm and an opposing force *F2* (in the –z direction of light propagation) at the other.

In normalized units these can respectively reach levels of 0.1 and 0.16 *P/c* (where *P* is the incident power per unit cell and *c* is the speed of light). In absolute terms, for an illumination intensity of 60 μW/μm$^2$, this corresponds to a force of ~20 fN on the cantilever tip and an opposing force of ~35 fN at the hinge, which would be sufficient to induce a static deformation (i.e. tilt) of only 2" - displacing the tip of the cantilever arm by ~5 pm. However, much larger deformations can be achieved by the same instantaneous driving forces at the structure's mechanical resonances, where displacement will be enhanced by the quality factor of the mechanical resonator. Assuming a Young's modulus of 150 GPa for the silicon membrane,[27] the first mechanical eigenmode of the 300 nm metamolecule cantilevers – the out-of-plane oscillation of the arms – is expected from numerical simulations to occur at a frequency of 165 MHz.

The optomechanical nonlinearity of the free-standing silicon membrane metamaterial was evaluated using the pump-probe experimental configuration illustrated schematically in Fig. 3a. Pump and probe beams at 1550 and 1310 nm respectively are generated by CW single-mode-fiber-coupled diode lasers, with the pump beam subsequently electro-optically modulated at frequencies up to 200 MHz. The beams are combined using a wavelength division multiplexer into a single fiber and then pass via a free-space collimator to the input port of an optical microscope operating in transmission mode. They are focused at normal incidence to concentric spots on the metamaterial, with diameters of ~10 μm. A fixed probe intensity of 25 μW/μm$^2$ is maintained at the sample, while peak pump intensity is varied up to a maximum level of 62.5 μW/μm$^2$. A low-pass filter blocks transmitted pump light and the probe signal is monitored using an InGaAs photodetector (New Focus 1811) connected to electrical network analyzer (Agilent Technologies E5071C). The sample is held under low vacuum conditions at ~0.1 mbar to reduce atmospheric damping of mechanical oscillations.

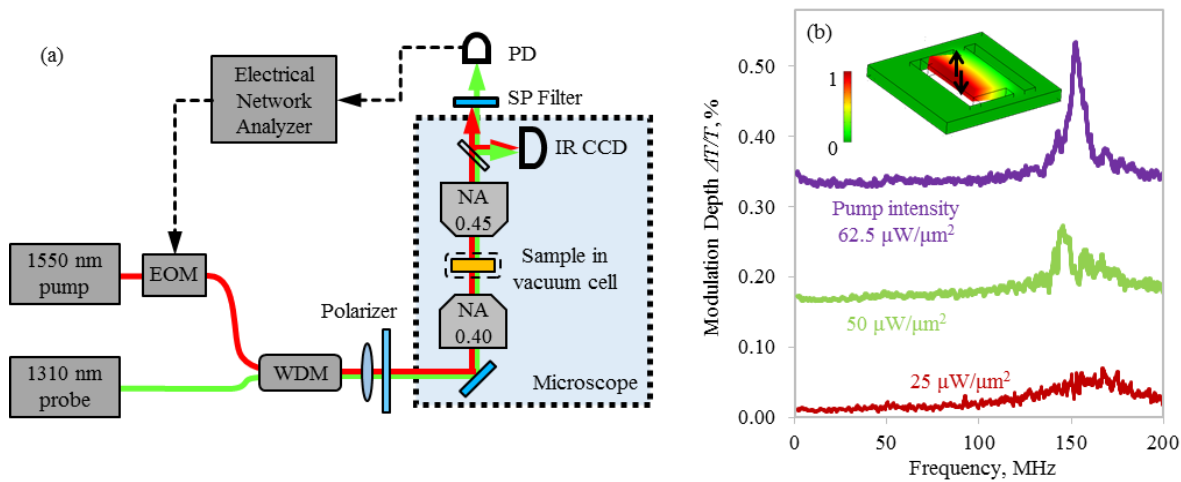

**Figure 3. Measuring the nonlinearity of nano-optomechanical all-dielectric metamaterials:** (a) Schematic of the pump-probe experimental arrangement for transmission-mode measurements of metamaterial nonlinear response. (b) 1310 nm [probe] transmission modulation depth as a function of pump [1550 nm] modulation frequency for a selection of peak pump intensities [as labelled]. The inset shows a nano-cantilever unit cell colored according to the relative magnitude of out-of-plane displacement, from numerical modelling, for the structure's first mechanical eigenmode.

Figure 3b presents the relative pump-induced change in probe transmission as a function of pump modulation frequency. As pump intensity increases the observed optomechanical resonance grows in strength and collapses spectrally to a central frequency of 152 MHz, reaching a maximum modulation depth of 0.2%. From numerical modelling, a transmission change of this magnitude corresponds to an induced nano-cantilever tilt of order 10', or a tip displacement of ~830 pm – some two orders of magnitude more than the expected static displacement at the same pump intensity. This implies a mechanical resonance quality factor of order 100, though we take this to be a lower limit on the value for individual silicon cantilevers inhomogeneously broadened (due to slight manufacturing defects and structural variations) across the metamaterial array. Indeed, at low pump intensities a spectrally disparate set of peaks emerges, suggestive of the distribution of individual cantilevers' different mechanical eigenfrequencies. At higher intensities the coupling among oscillators leads to synchronization and collective oscillation at a common frequency in good agreement with the computationally projected frequency of the structure's first mechanical eigenmode.

It is instructive to estimate what nonlinear susceptibility a hypothetical homogeneous medium would need to possess to provide a response of comparable magnitude to the nano-optomechanical silicon membrane metamaterial: Absorption in a nonlinear medium is conventionally described by the expression $-dI/dz = \alpha I + \beta I^2 + ...$, where $I$ is light intensity, $z$ is the propagation distance in the medium and $\alpha$ and $\beta$ are the linear and nonlinear absorption coefficients respectively. The observed nonlinear transmission change $\Delta T$ is proportional to the pump power, so can be quantified via an estimate of the first nonlinear absorption coefficient $\beta \sim \Delta T/(It)$, where $t$ is the metamaterial thickness. At the 152 MHz resonance frequency, $\beta \sim 7 \times 10^{-5}$ m/W, which corresponds to a nonlinear susceptibility of order $Im\{\chi^{(3)}\}/n^2 \sim 3.9 \times 10^{-14}$ m$^2$V$^{-2}$.

In conclusion, by structuring a free-standing nano-membrane of silicon at the sub-wavelength scale we engineer optical resonances strong enough to deliver a substantial optomechanical nonlinearity in an otherwise linear ultrathin medium. The nonlinear all-dielectric metamaterial operates at sub-GHz frequencies and μW/unit-cell intensities in the near-infrared spectral range. These free-standing all-dielectric metamaterials offer a compact, energy efficient and fast active optoelectronic platform potentially suited to practical application in high speed photonic applications. Improvements may be made in the design and fabrication of membrane metamaterials to enhance the probe transmission or reflectivity change per degree of tilt or nanometer of displacement, and to maximize the efficacy with which optical forces can generate such movements. But even while absolute changes are small, their sharply resonant nonlinear character may serve a variety of sensing (e.g. gas pressure, chemical binding) applications.

This work was supported by the UK Engineering and Physical Sciences Research Council [grant EP/G060363/1], the Samsung Advanced Institute of Technology [collaboration project number IO140325-01462-01], The Royal Society, the Singapore Ministry of Education [grant MOE2011-T3-1-005]. Following a period of embargo, the data from this paper can be obtained from the University of Southampton ePrints research repository, DOI: 10.5258 /SOTON/383566.